\documentstyle[psfig,mnras_cite,times]{mn}
\begin{document}
\author[Alan F. Heavens, Raul Jimenez and Ofer Lahav]{Alan F. Heavens$^{1}$, Raul Jimenez$^{1}$ and Ofer Lahav$^{2}$\\
$^{1}$ Institute for Astronomy, University of Edinburgh, Royal
Observatory, Blackford Hill, Edinburgh EH9 3HJ , United Kingdom\\
$^{2}$Institute of Astronomy, University of Cambridge, Madingley Road,
Cambridge CB3 0HA, U.K.}
\date{\today}
\title{Massive Lossless Data Compression and Multiple Parameter 
Estimation from Galaxy Spectra}
\newcommand{\be}{\begin{equation}}
\newcommand{\ee}{\end{equation}}
\newcommand{\ba}{\begin{eqnarray}}
\newcommand{\ea}{\end{eqnarray}}
\newcommand{\mat}{{\bf}}
\newcommand{\nn}{\nonumber\\}
\newcommand{\bA}{{\bf A}}
\newcommand{\bE}{{\bf E}}
\newcommand{\bB}{{\bf B}}
\newcommand{\bC}{{\bf C}}
\newcommand{\bF}{{\bf F}}
\newcommand{\bS}{{\bf S}}
\newcommand{\bfa}{{\bf a}}
\newcommand{\bb}{{\bf b}}
\newcommand{\bg}{{\bf g}}
\newcommand{\bj}{{\bf j}}
\newcommand{\bk}{{\bf k}}
\newcommand{\bn}{{\bf n}}
\newcommand{\bp}{{\bf p}}
\newcommand{\br}{{\bf r}}
\newcommand{\bu}{{\bf u}}
\newcommand{\bv}{{\bf v}}
\newcommand{\bx}{{\bf x}}
\newcommand{\by}{{\bf y}}
\newcommand{\bbmu}{\mbox{\boldmath $\mu$}}
\newcommand{\dd}{{\partial}}
\newcommand{\ddt}{{\partial\over \partial t}}
\def\gs{\mathrel{\raise1.16pt\hbox{$>$}\kern-7.0pt 
\lower3.06pt\hbox{{$\scriptstyle \sim$}}}}         
\def\ls{\mathrel{\raise1.16pt\hbox{$<$}\kern-7.0pt 
\lower3.06pt\hbox{{$\scriptstyle \sim$}}}}         

\maketitle

\begin{abstract}
We present a method for radical linear compression of datasets where
the data are dependent on some number $M$ of parameters.  We show
that, if the noise in the data is independent of the parameters, we
can form $M$ linear combinations of the data which contain as much
information about all the parameters as the entire dataset, in the
sense that the Fisher information matrices are identical; i.e. the
method is lossless.  We explore how these compressed numbers fare when
the noise is dependent on the parameters, and show that the method,
although not precisely lossless, increases errors by a very modest
factor.  The method is general, but we illustrate it with a problem
for which it is well-suited: galaxy spectra, whose data typically
consist of $\sim 10^3$ fluxes, and whose properties are set by a
handful of parameters such as age, brightness and a parametrised star
formation history.  The spectra are reduced to a small number of data,
which are connected to the physical processes entering the problem.
This data compression offers the possibility of a large increase in
the speed of determining physical parameters.  This is an important
consideration as datasets of galaxy spectra reach $10^6$ in size, and
the complexity of model spectra increases.  In addition to this
practical advantage, the compressed data may offer a classification
scheme for galaxy spectra which is based rather directly on physical
processes.
\end{abstract}

\section{Introduction}

There are many instances where objects consist of many data, whose
values are determined by a small number of parameters.  Often, it is
only these parameters which are of interest.  The aim of this paper is
to find linear combinations of the data which are focussed on
estimating the physical parameters with as small an error as possible.
Such a problem is very general, and has been attacked in the case of
parameter estimation in large-scale structure and the microwave
background (e.g. \pcite{TTH97} [hereafter TTH], \pcite{BJK98},
\pcite{Teg97a}, \pcite{Teg97b}).
Previous work has concentrated largely on the estimation of a single
parameter; the main advance of this paper is that it sets out a method
for the estimation of multiple parameters.  The method provides one
projection per parameter, with the consequent possibility of a massive
data compression factor.  Furthermore, if the noise in the data is
independent of the parameters, then the method is entirely lossless.
i.e. the compressed dataset contains as much information about the
parameters as the full dataset, in the sense that the Fisher
information matrix is the same for the compressed dataset as the
entire original dataset.  An equivalent statement is that the mean
likelihood surface is at the peak locally identical when the full or
compressed data are used. 

We illustrate the method with the case of galaxy spectra, for which
there are surveys underway which will provide $\sim 10^6$ objects.  In
this application, the noise is generally not independent of the
parameters, as there is a photon shot-noise component which depends on
how many photons are expected.  We take a spectrum with poor
signal-to-noise, whose noise is approximately from photon counting
alone, and investigate how the method fares.  In this case, the method
is not lossless, but the increase in error bars is shown to be
minimal, and superior in this respect to an alternative compression
system PCA (Principal Component Analysis).  

One advantage such radical compression offers is speed of analysis.
One major scientific goal of galaxy spectral surveys is to determine
physical parameters of the stellar component of the galaxies, such as
the age, star formation history, initial mass function and so on.
Such a process can, in principle, be achieved by generating model
galaxy spectra by stellar population synthesis techniques, and finding
the best-fitting model by maximum-likelihood techniques.  This can be
very time-consuming, and must inevitably be automated for so many
galaxies.  In addition, one may have a large parameter space to
explore, so any method which can speed up this process is worth
investigation.  One possible further application of the data
compression method is that the handful of numbers might provide the
basis of a classification scheme which is based on the physical
properties one wants to measure.

The outline of the paper is as follows: in section II we set out the
lossless compression method for noise which is independent of the
parameters; the proof appears in the appendix.  In section III we
discuss the more general case where the noise covariance matrix and
the mean signal both depend on the parameters.  In section IV we show
through a worked example of galaxy spectra that the method, although
not lossless, works very well in the general case. 

\section{Method}

We represent our data by a vector $\bx_i$, $i=1,\ldots,N$ (e.g. a set
of fluxes at different wavelengths).  These measurements include a
signal part, which we denote by $\bbmu$, and noise, $\bn$:
\begin{equation}
\bx = \bbmu + \bn
\end{equation}
Assuming the noise has zero mean, $\langle \bx\rangle = \bbmu$.  The
signal will depend on a set of parameters $\{\theta_\alpha\}$, which
we wish to determine.  For galaxy spectra, the parameters may be, for
example, age, magnitude of source, metallicity and some parameters
describing the star formation history.  Thus, $\bbmu$ is a noise-free
spectrum of a galaxy with certain age, metallicity etc.

The noise properties are described by the noise covariance matrix,
$\bC$, with components $C_{ij} = \langle n_i n_j\rangle$.  If the
noise is gaussian, the statistical properties of the data are
determined entirely by $\bbmu$ and $\bC$.  In principle, the noise can
also depend on the parameters.  For example, in galaxy spectra, one
component of the noise will come from photon counting statistics, and
the contribution of this to the noise will depend on the mean number
of photons expected from the source.

The aim is to derive the parameters from the data. If we assume
uniform priors for the parameters, then the a posteriori probability
for the parameters is the likelihood,  which for gaussian noise is
\ba
{\cal L}(\theta_\alpha) &=& {1\over (2\pi)^{N/2}
\sqrt{\det(\bC)}}
\times \nn & & 
\exp \left[-{1\over 2}\sum_{i,j} (x_i -
\mu_i)\bC^{-1}_{ij} (x_j-\mu_j)\right].
\ea
One approach is simply to find the (highest) peak in the likelihood,
by exploring all parameter space, and using all $N$ pixels.  The
position of the peak gives estimates of the parameters which are
asymptotically (low noise) the best unbiased estimators (see TTH).
This is therefore the best we can do.  The maximum-likelihood
procedure can, however, be time-consuming if $N$ is large, and the
parameter space is large.  The aim of this paper is to see whether we
can reduce the $N$ numbers to a smaller number, without increasing the
uncertainties on the derived parameters $\theta_\alpha$.  To be
specific, we try to find a number $N'<N$ of linear combinations of the
spectral data \bx\ which encompass as much as possible of the
information about the physical parameters.  We find that this can be
done losslessly in some circumstances; the spectra can be reduced to a
handful of numbers without loss of information.  The speed-up in
parameter estimation is about a factor $\sim 100$.

In general, reducing the dataset in this way will lead to larger error
bars in the parameters.  To assess how well the compression is doing,
consider the behaviour of the (logarithm of the) likelihood function
near the peak.  Performing a Taylor expansion and truncating at the
second-order terms,
\begin{equation}
\ln{\cal L} = \ln{\cal L}_{\rm peak} +{1\over 2}  {\partial^2 \ln {\cal L}\over
\partial \theta_\alpha \partial
\theta_\beta}\Delta\theta_\alpha\Delta\theta_\beta.
\end{equation}
Truncating here assumes that the likelihood surface itself is
adequately approximated by a gaussian everywhere, not just at the
maximum-likelihood point.   The actual likelihood surface will vary
when different data are used;  on average, though, the width is set by 
the (inverse of the) Fisher information matrix:
\begin{equation}
\bF_{\alpha\beta} \equiv - \left\langle {\partial^2 \ln {\cal L}\over
\partial \theta_\alpha \partial \theta_\beta} \right\rangle
\end{equation}
where the average is over an ensemble with the same parameters but different
noise.  

For a single parameter, the Fisher matrix $\bF$ is a scalar $F$,
and the error on the parameter can be no smaller than $F^{-1/2}$.  If
the data depend on more than one parameter, and all the parameters
have to be estimated from the data, then the error is larger.  The
error on one parameter $\alpha$ (marginalised over the others) is at
least $\left[(\bF^{-1})_{\alpha\alpha}\right]^{1/2}$ \scite{Kendall}.
There is a little more discussion of the Fisher matrix in
\scite{TTH97}, hereafter TTH.  The Fisher matrix depends on the signal 
and noise terms in the following way (TTH, equation 15)
\be
\bF_{\alpha\beta} = {1\over 2}{\rm
Tr}\left[\bC^{-1}\bC_{,\alpha}\bC^{-1} \bC_{,\beta}
+ \bC^{-1}(\bbmu_{,\alpha}\bbmu_{,\beta}^t + \bbmu_{,\beta}\bbmu_{,\alpha}^t)
\right].
\label{FisherFull}
\ee
where the comma indicates derivative with respect to the parameter.
If we use the full dataset $\bx$, then this Fisher matrix represents
the best that can possibly be done via likelihood methods with the
data.  

In practice, some of the data may tell us very little about the
parameters, either through being very noisy, or through having no
sensitivity to the parameters.  So in principle we may be able to
throw some data away without losing very much information about the
parameters.  Rather than throwing individual data away, we can do
better by forming linear combinations of the data, and then throwing
away the combinations which tell us least.  To proceed, we first
consider a single linear combination of the data:
\begin{equation}
y \equiv \bb^t \bx
\end{equation}
for some weighting vector \bb\ ($t$ indicates transpose).  We will try
to find a weighting which captures as much information about a
particular parameter, $\theta_\alpha$.  If we assume we know all the
other parameters, this amounts to maximising $\bF_{\alpha\alpha}$.  
The dataset (now consisting of a single number) has a Fisher matrix,
which is given in TTH (equation 25) by:
\begin{equation}
\bF_{\alpha\beta} = {1\over 2}\left({\bb^t \bC_{,\alpha}\bb\over  
\bb^t \bC\bb}\right)\left({\bb^t \bC_{,\beta}\bb\over  
\bb^t \bC\bb}\right) + {(\bb^t \bbmu_{,\alpha})(\bb^t \bbmu_{,\beta})
\over (\bb^t \bC\bb)}.
\label{Fisherb}
\end{equation}
Note that the denominators are simply numbers.  It is clear from this
expression that if we multiply $\bb$ by a constant, we get the same
$\bF$.  This makes sense: multiplying the data by a constant factor
does not change the information content.  We can therefore fix the
normalisation of $\bb$ at our convenience. To simplify the
denominators, we therefore maximise $\bF_{\alpha\alpha}$ subject to
the constraint
\begin{equation}
\bb^t \bC\bb = 1.
\end{equation}
The most general problem has both the mean ${\bf \bbmu}$ and the
covariance matrix \bC\ depending on the parameters of the spectrum,
and the resulting maximisation leads to an eigenvalue problem which is
nonlinear in \bb.  We are unable to solve this, so we consider a case
for which an analytic solution can be found.  TTH showed how to solve
for the case of estimation of a single parameter in two special cases:
1) when $\bbmu$ is known, and 2) when \bC\ is known (i.e. doesn't depend
on the parameters).  We will concentrate on the latter case,
but generalise to the problem of estimating many parameters at once.
For a single parameter, TTH showed that the entire dataset could be
reduced to a single number, with no loss of information about the
parameter.  We show below that, if we have $M$ parameters to estimate,
then we can reduce the dataset to $M$ numbers.  These $M$ numbers
contain just as much information as the original dataset; i.e. the
data compression is lossless.

We consider the parameters in turn.  With $\bC$ independent of the
parameters, $\bF$ simplifies, and, maximising $\bF_{11}$ subject to
the constraint requires
\begin{equation}
{\partial\over \partial b_i}\left(b_j \mu_{,1\, j}b_k 
\mu_{,1\, k} - \lambda b_j C_{jk} b_k\right) = 0
\end{equation}
where $\lambda$ is a Lagrange multiplier, and we assume the summation
convention ($j,k \in [1,N]$).  This leads to
\begin{equation}
\bbmu_{,1} (\bb^t\bbmu_{,1}) = \lambda \bC \bb
\end{equation}
with solution, properly normalised
\begin{equation}
\bb_1 = {\bC^{-1} \bbmu_{,1}\over \sqrt{\bbmu_{,1}^t
\bC^{-1}\bbmu_{,1}}}
\label{Evector1}
\end{equation}
and our compressed datum is the single number $y_1=\bb_1^t \bx$.  This
solution makes sense -- ignoring the unimportant denominator, the
method weights high those data which are parameter-sensitive, and low
those data which are noisy.  

To see whether the compression is lossless, we compare the Fisher
matrix element before and after the compression.  Substitution of
$\bb_1$ into (\ref{Fisherb}) gives
\be
\bF_{11} = \bbmu_{,1}^t \bC^{-1} \bbmu_{,1}
\label{FisherComp}
\ee
which is identical to the Fisher matrix element using the full data
(equation \ref{FisherFull}) if \bC\ is independent of $\theta_{1}$.
Hence, as claimed by TTH, the compression from the {\em entire}
dataset to the single number $y_1$ loses no information about
$\theta_{1}$.  For example, if $\bbmu \propto \theta$, then $y_1 =
\sum_i \bx_i/\sum_i \bbmu_i$ and is simply an estimate of the parameter
itself.

\subsubsection{Fiducial model}

It is important to note that $y_1$ contains as much information about
$\theta_1$ only if all other parameters are known, and also provided
that the covariance matrix and the derivative of the mean in
(\ref{Evector1}) are those at the maximum likelihood point.  We turn
to the first of these restrictions in the next section, and discuss
the second one here.  

In practice, one does not know beforehand what the true solution is,
so one has to make an initial guess for the parameters.  This guess we
refer to as the fiducial model.  We compute the covariance matrix
$\bC$ and the gradient of the mean ($\mu_{,\alpha}$) for this fiducial 
model, to construct $\bb_1$.   The Fisher matrix for the compressed
datum is (\ref{FisherComp}), but with the fiducial values inserted.
In general this is not the same as Fisher matrix at the true
solution.  In practice one can iterate: choose a fiducial model; use
it to estimate the parameters, and then repeat, using the estimate as
the estimated parameters as the fiducial model.  As our example in
section 4 shows, such iteration may be completely unnecessary.

\subsection{Estimation of many parameters}

The problem of estimating a single parameter from a set of data is
unusual in practice.  Normally one has several parameters to estimate
simultaneously, and this introduces substantial complications into the
analysis.  How can we generalise the single-parameter estimate above
to the case of many parameters?  We proceed by finding a second number
$y_2 \equiv \bb_2^t \bx$ by the following requirements:
\begin{itemize}
\item{$y_2$ is uncorrelated with $y_1$.  This demands that 
$\bb_2^t \bC \bb_1=0$.}
\item{$y_2$ captures as much information as possible about the
second parameter $\theta_{2}$.}
\end{itemize}
This requires two Lagrange multipliers (we normalise $\bb_2$ by demanding 
that $\bb_2^t \bC \bb_2 = 1$ as before).  Maximising and applying the
constraints gives the solution
\begin{equation}
\bb_2 = {\bC^{-1}\bbmu_{,2} - (\bbmu_{,2}^t \bb_1)\bb_1 \over
\sqrt{\bbmu_{,2} \bC^{-1} \bbmu_{,2} - (\bbmu_{,2}^t
\bb_1)^2}}.
\label{bb2}
\end{equation}
This is readily generalised to any number $M$ of parameters.  There
are then $M$ orthogonal vectors $\bb_m$, $m=1, \ldots M$, each $y_m$
capturing as much information about parameter $\alpha_m$ which is not
already contained in $y_q;\ q<m$.  The constrained maximisation gives
\begin{equation}
\bb_m = {\bC^{-1}\bbmu_{,m} - \sum_{q=1}^{m-1}(\bbmu_{,m}^t 
\bb_q)\bb_q \over
\sqrt{\bbmu_{,m} \bC^{-1} \bbmu_{,m} - \sum_{q=1}^{m-1}
(\bbmu_{,m}^t \bb_q)^2}}.
\label{bbm}
\end{equation} 
This procedure is analogous to Gram-Schmidt orthogonalisation with a
curved metric, with $\bC$ playing the role of the metric tensor.  Note
that the procedure gives precisely $M$ eigenvectors and hence $M$
numbers, so the dataset has been compressed from the original $N$ data
down to the number of parameters $M$.

Since, by construction, the numbers $y_m$ are uncorrelated, the
likelihood of the parameters is obtained by multiplication of the
likelihoods obtained from each statistic $y_m$.  The $y_m$ have mean
$\langle y_m \rangle = \bb_m^t \bbmu$ and unit variance, so the
likelihood from the compressed data is simply 
\be
\ln{\cal L}(\theta_\alpha) = {\rm constant} - \sum_{m=1}^{M}
{(y_m-\langle y_m\rangle)^2\over 2}
\ee
and the Fisher matrix of the combined numbers is just the sum of the
individual Fisher matrices.  Note once again the role of the fiducial
model in setting the weightings $\bb_m$: the orthonormality of the new
numbers only holds if the fiducial model is correct.  Multiplication
of the likelihoods is thus only approximately correct, but iteration
could be used if desired.

\subsubsection{Proof that the method can be lossless for 
many parameters}

Under the assumption that the covariance matrix is independent of the
parameters, reduction of the original data to the $M$ numbers $y_m$
results in no loss of information about the $M$ parameters at all.  In
fact the set $\left\{y_m\right\}$ produces, on average, a likelihood
surface which is locally identical to that from the entire dataset --
no information about the parameters is lost in the compression
process.  With the restriction that the information is defined locally
by the Fisher matrix, the set $\{y_m\}$ is a set of sufficient
statistics for the parameters $\{\theta_\alpha\}$ (e.g. \pcite{Koch}).
A proof of this for an arbitrary number of parameters is given in the
appendix.

\section{The general case}

In general, the covariance matrix does depend on the parameters, and
this is the case for galaxy spectra, where at least one component of
the noise is parameter-dependent.  This is the photon counting noise,
for which $\bC_{ii} = \bbmu_i$.  TTH argued that it is better to treat
this case by using the $n$ eigenvectors which arise from assuming the
mean is known, rather than the single number (for one parameter) which
arises if we assume that the covariance matrix is known, as above.  We
find that, on the contrary, the small number of eigenvectors $\bb_m$
allow a much greater degree of compression than the known-mean
eigenvectors (which in this case are simply individual pixels, ordered by
$|\bbmu_{,\alpha}/\bbmu|$).  For data signal-to-noise of around 2, the latter
allow a data compression by about a factor of 2 before the errors on
the parameters increase substantially, whereas the method here allows
drastic compression from thousands of numbers to a handful.  To show
what can be achieved, we use a set of simulated galaxy spectra to
constrain a few parameters characterising the galaxy star formation
history.

\subsection{Parameter Eigenvectors}

In the case when the covariance matrix is independent of the
parameters, it does not matter which parameter we choose to form
$y_1$, $y_2$, etc, as the likelihood surface from the compressed
numbers is, on average, locally identical to that from the full
dataset.  However, in the general case, the procedure does lose
information, and the amount of information lost could depend on the
order of assignment of parameters to $m$.  If the parameter estimates
are correlated, as we will see in Fig. \ref{LikeFull}, the error
in both parameters is dominated by the length of the likelihood
contours along the `ridge'.  It makes sense then to diagonalise the
matrix of second derivatives of $\ln{\cal L}$ at the fiducial model, and use
these as the parameters (temporarily), as proposed by
\pcite{Ballinger00} for galaxy surveys.  The parameter eigenvalues
would order the importance of the parameter combinations to the
likelihood.  The procedure would be to take the smallest eigenvalue
(with eigenvector lying along the ridge), and make the likelihood
surface as narrow as possible in that direction.  One then repeats
along the parameter eigenvectors in increasing order of eigenvalue.

Specifically, diagonalise $\bF_{\alpha\beta}$ in (\ref{FisherFull}),
to form a diagonal covariance matrix $\Lambda = \bS^t \bF \bS$.  The
orthogonal parameter combinations are $\psi = \bS^t \theta$, where
\bS\ has the normalised eigenvectors of \bF\ as its columns.  The
weighting vectors $\bb_m$ are then computed from (\ref{bbm}) by
replacing $\bbmu_{,\alpha p}$ by $\bS_{p r} \bbmu_{,\alpha r}$.

\section{A Worked Example: galaxy spectra}

\begin{figure}
\centerline{
\psfig{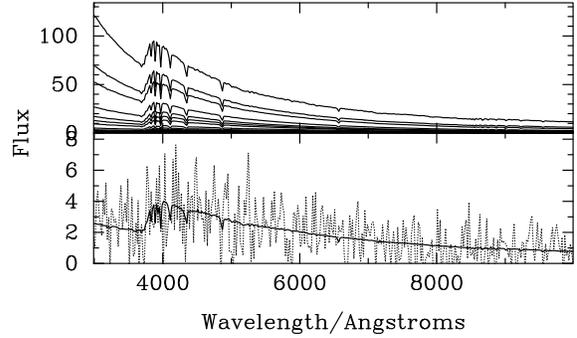}}
\caption[]{\label{Spectra} Top panel: example model spectra, with age 
increasing downwards. Bottom panel: simulated galaxy spectrum
(including noise), whose properties we wish to determine, superimposed
on noise-free spectrum of galaxy with the same age.}
\end{figure}

We start by investigating a two-parameter model.  We have run a grid
of stellar evolution models, with a burst of star formation at time
$-t$, where $t=0$ is the present day.  The star formation rate is
$SFR(t') = A \delta(t'+t)$ where $\delta$ is a Dirac delta function.
The two parameters to determine are age $t$ and normalisation $A$.
Fig. \ref{Spectra} shows some spectra with fixed normalisation ($1
M_\odot$ of stars produced) and different age.  There are $n=352$
pixels between 300 and 1000 nm.  Real data will be more complicated
(variable transmission, instrumental noise etc) but this system is
sufficiently complex to test the methods in essential respects.  For
simplicity, we assume that the noise is gaussian, with a variance
given by the mean, $\bC = {\rm diag}(\bbmu_1, \ldots)$.  This is
appropriate for photon number counts when the number is large.  We
assume the same behaviour, even with small numbers, for illustration,
but there is no reason why a more complicated noise model cannot be
treated.  It should be stressed that this is a more severe test of the
model than a typical galaxy spectrum, where the noise is likely to be
dominated by sources independent of the galaxy, such as CCD read-out
noise or sky background counts.  In the latter case, the compression
method will do even better than the example here.
%
%

The simulated galaxy spectrum is one of the galaxy spectra (age 3.95
Gyr, model number 100), and the maximum signal-to-noise per bin is taken to be
2.  Noise is added, approximately photon noise, with a gaussian
distribution with variance equal to the number of photons in each
channel (Fig. \ref{Spectra}).  Hence $\bC =$ diag($\bbmu_1,\bbmu_2,\ldots)$.
%
%

\begin{figure}
\centerline{
\psfig{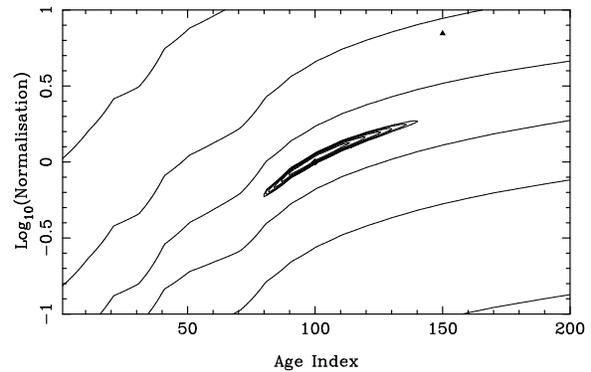}}
\caption[]{\label{LikeFull} Full likelihood solution using all pixels.
There are 6 contours running down from the peak value in steps of 0.5 
(in $\ln{\cal L}$), and 3 outer contours at $-100$, $-1000$ and $-10000$.
The triangle in the upper-right corner marks the fiducial model which
determines the eigenvectors $\bb_{1,2}$.}
\end{figure}
\begin{figure}
\centerline{
\psfig{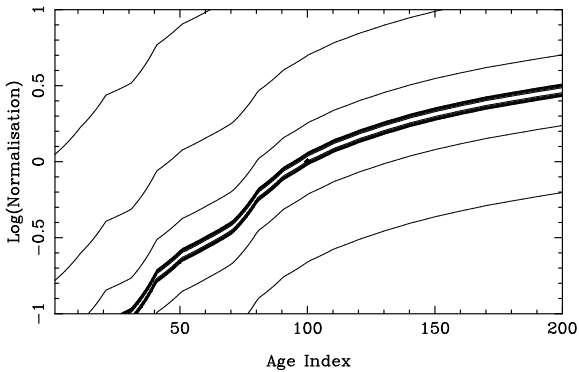}}
\caption[]{\label{LikeAge}  Likelihood solution for the age datum
$y_1$.  Contours are as in Fig. \ref{LikeFull}.}
\end{figure}
\begin{figure}
\centerline{
\psfig{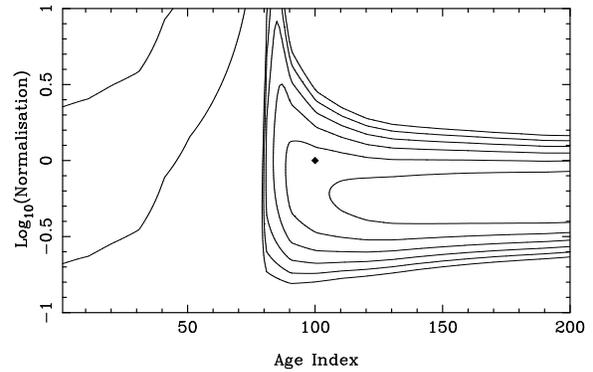}}
\caption[]{\label{LikeNorm}  Likelihood solution for the normalisation datum
$y_2$.  Contours are as in Fig. \ref{LikeFull}.}
\end{figure}
\begin{figure}
\centerline{
\psfig{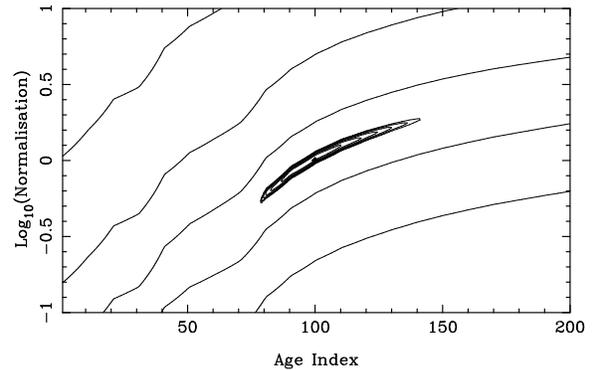}}
\caption[]{\label{LikeBoth}  Likelihood solution for the age datum
$y_1$ and the normalisation datum $y_2$.  Contours are as in
Fig. \ref{LikeFull}.}
\end{figure}

The most probable values for the age and normalisation (assuming
uniform priors) is given by maximising the likelihood:
\ba
{\cal L}({\rm age,norm}) &=& {1\over (2\pi)^{n/2}
\sqrt{\prod_i \mu_i}}
\times \nn & & 
\exp \left[-{1\over 2}\sum_i (x_i - \mu_i)^2/\mu_i \right]
\ea
where $\bbmu$ depends on age and normalisation.  $\ln{\cal
L}$ is shown in Fig. \ref{LikeFull}.  Since this uses all the data,
and all the approximations hold, this is the best that can be done,
given the S/N of the spectrum.
%
%
\begin{figure}
\centerline{
\psfig{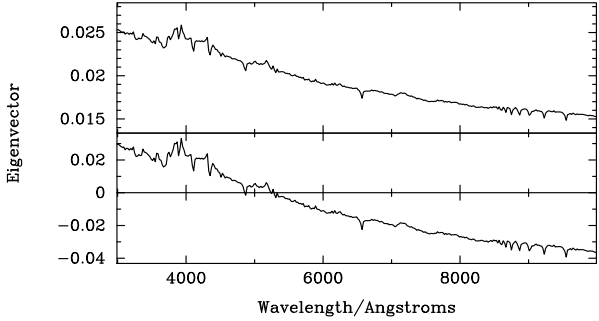}}
\caption[]{\label{Evectors} Eigenvectors $-\bb_1$ (age) and
 $-\bb_2$ (normalisation).  Wavelength $\lambda$ is in Angstroms.  Note
that the weights in $\bb_1$ are negative, which is why the sign has
been changed for plotting: the blue (left) end of the spectrum which
is weighted most heavily for $y_1$.  This is expected as this part of
the spectrum changes most rapidly with age.  Note that these
weightings differ by a constant;  this feature is special to the
amplitude parameter, and is explained in the text.}
\end{figure}
To solve the eigenvalue problem for \bb\ requires an initial guess for
the spectrum.  This `fiducial model' was chosen to have an age of 8.98
Gyr, i.e. very different from the true solution (model number 150
rather than 100).  This allows us to compute the eigenvector $\bb_1$
from (\ref{Evector1}).  This gives the single number $y_1 = \bb_1^t
\bx$.  With this as the datum, the likelihood for age and
normalisation is
\be
{\cal L}({\rm age,norm})  =  {1\over \sqrt{2\pi}}\exp \left[-{(y_1 -
\langle y_1\rangle)^2\over 2}\right]
\end{equation}
where $\langle y_1\rangle = \bb_1^t \bbmu$.  Note that the mean and
covariance matrix here depend on the parameters - i.e. they are not
from the fiducial model. The resultant likelihood is shown in
Fig. \ref{LikeAge}.  Clearly it does less well than the full solution,
but it does constrain the parameters to a narrow ridge, on which the
true solution (age model=100, $\log({\rm normalisation})=0$ lies.
%
%
The second eigenvector $\bb_2$ is obtained by taking the normalisation
as the second parameter.  The vector is shown in the lower panel of
Fig. \ref{Evectors}.  The normalisation parameter is rather a special
case, which results in $\bb_2$ differing from $\bb_1$ only by a
constant offset in the weights  (For this parameter $\bbmu_{,\alpha} = 
\bbmu$ and so $\bC^{-1}\bbmu_{,\alpha} = (1,1,\ldots,1)^t)$.
The likelihood for the parameters with $y_2$ as the single datum is
shown in Fig. \ref{LikeNorm}.  On its own, it does not tightly
constrain the parameters, but when combined with $y_1$, it does remarkably
well (Fig. \ref{LikeBoth}).
%
%
%
%

\subsection{Three-parameter estimation}

\begin{figure}
\centerline{
\psfig{figure=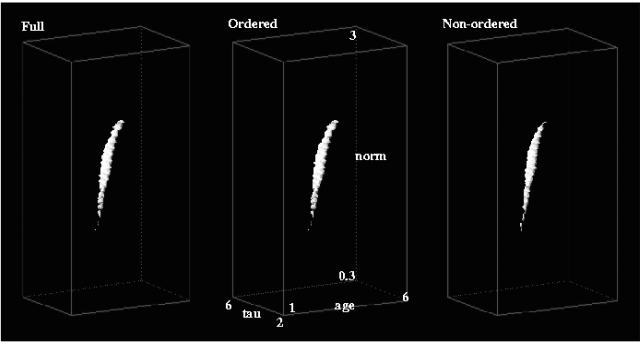,width=8.5cm,angle=0}}
\caption[]{\label{3D}  (Left) Likelihood solution for the full
dataset of 1000 numbers for a single galaxy, as a function of $t$/Gyr,
$\tau$/Gyr and amplitude.  (Middle) Likelihood for 3 compressed
numbers, from parameter eigenvectors.  (Right) likelihood surface from
3 compressed numbers (age, normalisation and $\tau$ eigenvectors). All
contours shown are 3.13 below the peak in $\ln{\cal L}$; the
irregularities in the surface are artefacts of the surface-drawing
routine.}
\end{figure}

We complicate the situation now to a 3-parameter star-formation rate $SFR(t)=A
\exp(-t/\tau)$,  and estimate $A$, $t$ and $\tau$.  Chemical
evolution is included by using a simple closed-box model (with
instantaneous recycling; \pcite{Pagel97}).  This
affects the depths of the absorption lines.  If we follow the same
procedure as before, choosing ($t,A,\tau$) as the order for computing
$\bb_1$, $\bb_2$ and $\bb_3$, then the product of the likelihoods from
$y_1$, $y_2$ and $y_3$ is as shown in the right panel of
Fig. \ref{3D}.  The left panel shows the likelihood from the full
dataset of 1000 numbers, which does little better than the 3
compressed numbers.  It is interesting to explore how the parameter
eigenvector method fares in this case.  Here we follow the procedure
in section 2, and maximise the curvature along the ridge first.  The
resulting three numbers constrain the parameters as in the middle
panel; in this case there is no apparent improvement over using
eigenvectors from $(t,A,\tau)$, but it may be advantageous in other
applications.
%
%

\subsection{Estimate of increase in errors}

For the noise model we have adopted, we can readily compute the
increase in the conditional error for one of the parameters - the
normalisation of the spectrum.  This serves as an illustration of how
much information is lost in the compression process.  In this case,
$\bC=\bbmu$, and $\bC_{,\alpha}=\bbmu_{,\alpha}=\bbmu_{\alpha}$, and the
Fisher matrix (a single element) can be written in terms of the total
number of photons and the number of spectral pixels.  From (\ref{FisherFull}),
$F^O=N_{\rm photons} + N_{\rm pixels}/2$.   The compressed data, on
the other hand, have a Fisher matrix $F=N_{\rm photons}+1/2$, so the
error bar on the normalisation is increased by a factor
\be
{\rm Fractional\ error\ increase} \simeq \sqrt{1+{1\over 2s}}
\ee
for $N_{\rm photons}\gg 1$, and $s\equiv N_{\rm photons}/N_{\rm
pixels}$ is the average number of photons per pixel.  Even if $s$ is
as low as 2, we see that the error bar is increased only by around
12\%.

\subsection{Computational issues}

We have reduced the likelihood problem in this case by a factor of
more than a hundred.  The eigenproblem is trivial to solve.  The work
to be done is in reducing a whole suite of model spectra to $M$
numbers, and by forming scalar products of them with the vectors $\bb_m$.
This is a one-shot task, and trivial in comparison with the job of
generating the models.  

\subsection{Role of fiducial model}

\begin{figure}
\centerline{
\psfig{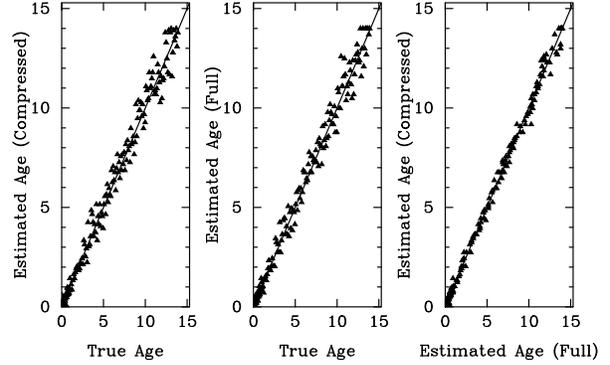}}
\caption[]{\label{SN2F} The effect of the fiducial model on recovery
of the parameters.  Here a single fiducial model is chosen (with age 9 
Gyr), and ages recovered from many true galaxy spectra with ages
between zero and 14 Gyr. Top left panel shows the recovered age from the
two numbers $y_1$ and $y_2$ (with age and normalisation weightings),
plotted against the true model age.  Top right shows how well the 
full dataset (with S/N $\ls 2$) can recover the parameters.  The lower
panel shows the estimated age from the $y_1$ and $y_2$ plotted against 
the age recovered from the full dataset, showing that the compression
adds very little to the error, even if the fiducial model is very
wrong.  Note also that the scatter increases with age;  old galaxies
are more difficult to date accurately.}
\end{figure}

The fiducial model sets the weightings $\bb_m$.  After this step, the
likelihood analysis is correct for each $y_m$, even if the fiducial
model is wrong.  The only place where there is an approximation is in
the multiplication of the likelihoods from all $y_m$ to estimate
finally the parameters.  The $y_m$ are strictly only uncorrelated if
the fiducial model coincides with the true model.  This approximation
can be dropped, if desired, by computing the correlations of the $y_m$
for each model tested.  We have explored how the fiducial model
affects the recovered parameters, and an example result from the
two-parameter problem is shown in Fig. \ref{SN2F}.  Here the ages and
normalisations of a set of `true' galaxies with S/N $\ls 2$ are
estimated, using a common (9Gyr) galaxy as the fiducial model. We see
that the method is successful at recovering the age, even if the
fiducial model is very badly wrong.  There are errors, of course, but
the important aspect is whether the compressed data do significantly
worse than the full dataset of 352 numbers.  Fig. \ref{SN2F} shows
that this is not the case.

Although it appears from this example to be unnecessary, if one wants
to improve the solution, then it is permissible to iterate, using the first
estimate as the fiducial model.  This adds to the computational task,
but not significantly;  assuming the first iteration gives a
reasonable parameter estimate, then one does not have to explore the
entire parameter space in subsequent iterations.
%
%

\section{Comparison with Principal Component Analysis}

It is interesting to compare with other data compression and parameter
estimation methods.  For example, Principal Component Analysis is
another linear method (e.g. \pcite{Murtagh87}, \pcite{Francis92},
\pcite{Connolly95},
\pcite{Folkes96},\pcite{Sodre96},
\pcite{Galaz98}, \pcite{Bromley98}, \pcite{Glazebrook98},
\pcite{Singh98}, \pcite{Connolly99}, \pcite{Ronen99},
\pcite{Folkes99}), which projects the data onto eigenvectors of
the covariance matrix, which is determined empirically from the
scatter between flux measurements of different galaxies.  Part of the
covariance matrix in PCA is therefore determined by differences in the 
models, whereas in our case $\bC$ refers to the noise alone.  PCA then 
finds uncorrelated projections which contribute in decreasing amounts
to the variance between galaxies in the sample.    

\begin{figure}
\centerline{
\psfig{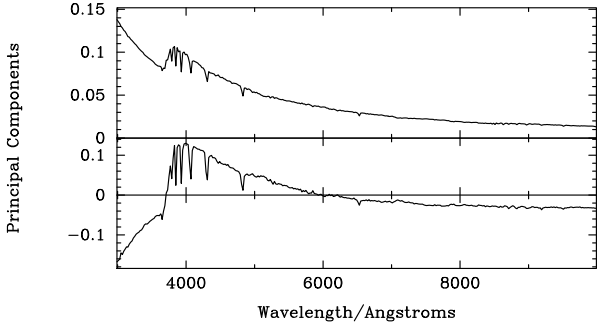}}
\caption[]{\label{PC}  The first two principal component eigenvectors,
from a system of model spectra consisting of a burst at different times.}
\end{figure}

\begin{figure}
\centerline{
\psfig{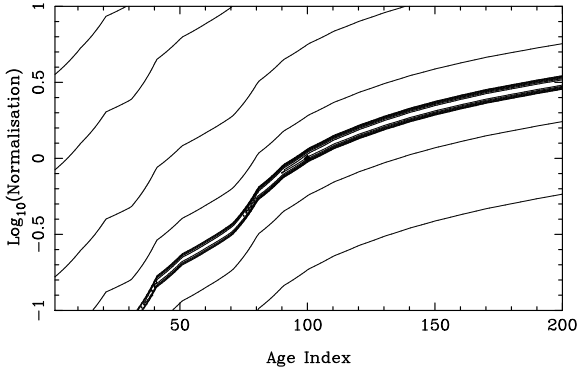}}
\caption[]{\label{LikePCA} Likelihood solution for the first two
principal components, PC1 (top) and PC2.  Contours are as in
Fig. \ref{LikeFull}.}
\end{figure}

\begin{figure}
\centerline{
\psfig{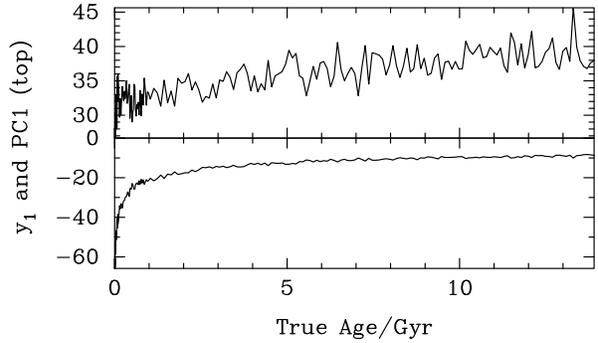}}
\caption[]{\label{PC1y1} First principal component (PC1) and $y_1$ versus
age.  One in every 10 models was used to do the PCA.  In the presence
of noise, at a level of $S/N<2$ per bin, $y_1$ is almost monotonic
with age, whereas PC1, although correlated with age, is not a good age
estimator.}
\end{figure}

One finds that the first principal component is correlated with the
galaxy age \cite{Ronen99}.  Figure \ref{PC} shows the PCA eigenvectors
obtained from a set of 20 burst model galaxies which differ only in
age, and Figure \ref{LikePCA} shows the resultant likelihood from the
first two principal components.  In the language of this paper, the
principal components are correlated, so the $2\times 2$ covariance
matrix is used to determine the likelihood.  We see that the
components do not do nearly as well as the parameter eigenvectors;
they do about as well as $y_1$ on its own.  For interest, we plot the
first principal component and $y_1$ vs. age in Figure \ref{PC1y1}.  In
the presence of noise ($S/N < 2$ per bin), $y_1$ is almost monotonic
with age, whereas PC1 is not.  Since PCA is not optimised for
parameter estimation, it is not lossless, and it should be no surprise
that it fares less well than the tailored eigenfunctions of section
III.  If one cannot model the effect of the parameters a priori, then
this method cannot be used,  whereas PCA might still be an effective tool.
%
%
%
%
%
%

\section{Discussion}

We have presented a linear data compression algorithm for estimation
of multiple parameters from an arbitrary dataset.  If there are $M$
parameters, the method reduces the data to a compressed dataset with
$M$ members.  In the case where the noise is independent of the
parameters, the compression is lossless; i.e. the $M$ data contain as
much information about the parameters as the entire dataset.
Specifically, this means the mean likelihood surface around the peak
is locally identical whichever of the full or compressed dataset is
used as the data.  It is worth emphasising the power of this method:
it is well known that, in the low-noise limit, the maximum likelihood
parameter esimates are the best unbiased estimates.  Hence if we do as 
well with the compressed dataset as with the full dataset, there is no 
other method, linear or otherwise, which can improve upon our results.
The method can result in a massive compression, with the degree of
compression given by the ratio of the size of the dataset to the
number of parameters.  Parameter estimation is speeded up by the same
factor.

Although the method is lossless in certain circumstances, we believe
that the data compression can still be very effective when the noise
does depend on the model parameters.  We have illustrated this using
simulated galaxy spectra as the data, where the noise comes from
photon counting (in practice, other sources of noise will also be
present, and possibly dominant); we find that the algorithm is still
almost lossless, with errors on the parameters increasing typically by
a factor $\sim \sqrt{1+1/(2s)}$, where $s$ is the average number of
photons per spectral channel.  The example we have chosen is a more
severe test of the algorithm than real galaxy spectra; in reality the
noise may well be dominated by factors external to the galaxy, such as
detector read-out noise, sky background counts (for ground-based
measurements) or zodiacal light counts (for space telescopes).  In
this case, the noise is indeed independent of the galaxy parameters,
and the method is lossless.   

The compression method requires prior choice of a fiducial model,
which determines the projection vectors $\bb$.  The choice of fiducial
model will not bias the solution, and the likelihood given the $y_m$
individually can be computed without approximation.  Combining the
likelihoods by multiplication from the individual $y_m$ is
approximate, as their independence is only guaranteed if the fiducial
model is correct.  However, in our examples, we find that the method
correctly recovers the true solution, even if the fiducial model is
very different.  If one is cautious, one could always iterate.  There
are circumstances where the choice of a good fiducial model may be
more important, if the eigenvectors depend very sensitively on the
model parameters.  An example of this is the determination of the
redshift $z$ of the galaxy, whose observed wavelengths are increased
by a factor $1+z$ by the expansion of the Universe.  If the main
signal for $z$ comes from spectral lines, then the method will give
great weight to certain discrete wavelengths, determined by the
fiducial $z$. If the true redshift is different, these wavelengths
will not coincide with the spectral lines.  It should be stressed that
the method will still allow an estimate of the parameters, including
$z$, but the error bars will not be optimal.  This may be one case
where applying the method iteratively may be of great value.

We have compared the parameter estimation method with another linear
compression algorithm, Principal Component Analysis.  PCA is not
lossless unless all principal components are used, and compares
unfavourably in this respect for parameter estimation.  However, one
requires a theoretical model for the methods in this paper; PCA does
not require one, needing instead a representative ensemble for
effective use.  Other, more ad hoc, schemes consider particular
features in the spectrum, such as broad-band colours, or equivalent
widths of lines \cite{Worthey94}.  Each of these is a ratio of linear
projections, with weightings given by the filter response or sharp
filters concentrated at the line.  There may well be merit in the way
the weightings are constructed, but they will not in general do as
well as the optimum weightings presented here.  It is worth remarking
on the ability of the method to separate parameters such as age and
metallicity, which often appear degenerately in some methods.  In the
`external noise' case, then {\em provided} the degeneracy can be
lifted by maximum likelihood methods using every pixel in the
spectrum, then it can also be lifted by using the reduced data.  Of
course, if the modelling is not adequate to estimate the parameters
using all the data, then compression is not going to help at all, and
one needs to think again.  For example, a complication which may arise
in a real galaxy spectrum is the presence of features not in the
model, such as emission lines from hot gas.  These can be included if
the model is extended by inclusion of extra parameters.  This problem
exists whether the full or compressed data are used.  Of course, we
can use standard goodness-of-fit tests to determine whether the data
are consistent with the model as specified, or whether more parameters
are required.

The data compression to a handful of numbers offers the possibility of
a classification scheme for galaxy spectra.  This is attractive as the
numbers are connected closely with the physical processes which
determine the spectrum, and will be explored in a later paper.  An
additional realistic aim is to determine the star formation history of
each individual galaxy, without making specific assumptions about the
form of the star formation rate.  The method in this paper provides
the means to achieve this.

\noindent{\bf Acknowledgments}

\noindent We thank Andy Taylor and Rachel Somerville, and the referee, 
Paul Francis, for useful comments.  Computations were made using
Starlink facilities.

\vspace{1cm}

\noindent{\bf Appendix}

In this appendix, we prove that the linear compression algorithm for
estimation of an arbitrary number $M$ of parameters is lossless,
provided the noise is independent of the parameters,
$\bC_{,\alpha}=0$.  Specifically, loss-free means the Fisher matrix
for the set of $M$ numbers $y_m=\bb_m^t \bx$ is identical to the
Fisher matrix of the original dataset $\bx$:
\be 
F_{\alpha\beta}^O = \langle\alpha|\beta\rangle.  
\ee 
By construction, the $y_m$ are uncorrelated, so the likelihoods
multiply and the Fisher matrix for the set $\{y_m\}$ is the sum of the
derivatives of the log-likelihoods from the individual $y_m$:
\be
F_{\alpha\beta}=\sum_m F_{\alpha\beta}(m).
\ee
From (\ref{Fisherb}),
\be
F_{\alpha\beta}(m)=(\bb_m^t \bbmu_{,\alpha})(\bb_m^t \bbmu_{,\beta})
\ee
With (\ref{bbm}), we can write
\be
\bb_m^t ={\bbmu_{,m}^t \bC^{-1} - \sum_{q=1}^{m-1}(\bb_q^t
\bbmu_{,m}^t)\bb_q \over
\sqrt{\langle m| m\rangle -  \sum_{q=1}^{m-1}(\bb_q^t
\bbmu_{,m}^t)^2}}.
\ee
Hence 
\ba
F_{\alpha\beta}(m)& =&\left[\langle \alpha | m \rangle - \sum_{q=1}^{m-1}
F_{\alpha m}(q)\right]\times \nn
& & {\left[\langle \beta | m \rangle - \sum_{q=1}^{m-1}
F_{\beta m}(q)\right]\over \left[\langle m | m \rangle - \sum_{q=1}^{m-1}
F_{m m}(q)\right]}\nn
\label{Fabm}
\ea
Consider first $\beta=m$:
\begin{eqnarray}
F_{\alpha m}(m)&=&\langle \alpha | m \rangle -
\sum_{q=1}^{m-1}F_{\alpha m}(q)\nn
\Rightarrow F_{\alpha M} &=& \sum_{q=1}^{M}F_{\alpha M}(q) = \langle
\alpha | M \rangle =  F_{\alpha M}^O\nn
\label{Fam}
\end{eqnarray}
proving that these terms are unchanged after compression.  We
therefore need to consider $F_{\alpha\beta}(m)$ for $\alpha$ or
$\beta<m$.  First we note that 
\be 
F_{\alpha\beta}(m)={F_{\alpha m}(m)F_{m \beta}(m)\over F_{m m}(m)}
\label{FFF}
\ee
and, from (\ref{Fam}),
\be
\sum_{q=1}^{\beta}F_{\alpha \beta}(q) = \langle \alpha | \beta \rangle 
\label{sum}
\ee
We want the sum to extend to $M$.  However, the terms from $\beta+1$
to $M$ are all zero.  This can be shown as follows: (\ref{FFF}) shows
that it is sufficient to show that $F_{\alpha m}(m)=0$ if $m>\alpha$.
Setting $\beta=m$ in (\ref{sum}), and reversing $\alpha$ and $m$, we
get
\be
\sum_{\alpha+1}^{m}F_{\alpha m}(q) = 0.
\label{sum2}
\ee
Now, the contribution from $q$ does not depend on derivatives wrt
higher-numbered parameters, so we can evaluate $F_{\alpha
m}(\alpha+1)$ by setting $m=\alpha+1$.  The sum (\ref{sum2}) implies
that this term zero.  Increasing $m$ successively by one up to $M$,
and using (\ref{sum2}), proves that all the terms are zero, proving
that the compression is lossless.


\end{document}